\documentclass[preprint,showpacs,preprintnumbers,amsmath,amssymb]{revtex4}

\usepackage{graphicx}
\usepackage{dcolumn}
\usepackage{bm}

\begin{document}


\title{Conductivity and Spin Susceptibility for the Disordered 2D Hubbard Model}

\author{Kohjiro Kobayashi}
\affiliation{%
Department of Physics, Ohio State University, Columbus, OH 43210\\
}%
\author{Byounghak Lee}
\affiliation{%
Computational Research Division, Lawrence Berkeley National Laboratory, Berkeley, California 94720\\
}%
\author{Nandini Trivedi}
\affiliation{%
Department of Physics, Ohio State University, Columbus, OH 43210\\
}%

\date{\today}

\begin{abstract}
The effect of disorder on a class of transition metal oxides described by a single orbital Hubbard model at half filling is investigated. The phases are characterized by the nature of the electronic and spin excitations. The frequency and temperature-dependent conductivity and spin susceptibility as functions of disorder are calculated. The interplay of disorder and electron-electron interaction produces unusual behavior in this system. For example, the dc conductivity, which is vanishingly small at low disorder in the Mott phase and at high disorder in the localized phase, gets surprisingly enhanced at intermediate disorder in a "metallic" phase. Moreover, the spin susceptibility in this "metallic" phase is not the expected Pauli-behavior but Curie-$1/T$ due to the presence of local moments.
\end{abstract}

\maketitle

\section{Introduction}

Based on the scaling theory, it is expected that non-interacting electrons are localized for any weak disorder in two dimensions in zero magnetic field \cite{abrahams1979}. This theoretical result was questioned when transport experiments in 2D silicon metal-oxide-semiconductor field-effect transistors (MOSFET's) showed a metallic phase with $\frac{d\rho}{dT}>0$, where $\rho$ is the resistivity, at higher density of carriers than the critical density of carriers, $n_c$ and an insulating phase with $\frac{d\rho}{dT}<0$ at lower density of carriers, $n<n_c$ \cite{kravchenko1994}. Further experiments supported the observation of a MIT in 2D systems \cite{popovic1997, simonian1997,coleridge1997,lam1997,hanein1998,simmons1998,mills1999,papadakis1999}.

The interplay of disorder and electron-electron interaction can have an important effect in 2D systems. The recent experiments have shown that electron-electron interactions can lead to a possible metallic phase in the 2D electron gas. Motivated by these experiments, we investigate the 2D disordered Hubbard model at half-filling \cite{heidarian2004}. In previous work, a possible metallic phase at $T=0$ with extended wave functions, sandwiched between a Mott insulator for low disorder and a localized insulator for large disorder, is found. In the metallic phase, the spectral gap is zero but antiferromagnetism is found to persist. The existence of a metallic phase is confirmed by the finite size scaling behavior of the inverse participation ratio (IPR). At intermediate disorder, the IPR for $U=0$ extrapolates to a finite value as $L\rightarrow\infty$. Thus, the system is a localized. However, for the interacting system, $U=4t$, the IPR as a function of $1/L$ extrapolates to zero in the thermodynamic limit, implying a diverging localization length would occur in a metallic phase. The fact that the non-interacting system with the same disorder realization had a finite localization length indicates that interactions clearly played an important role in delocalizing the wave functions or at least in making them much more extended than the wave functions in the absence of interaction.

\begin{figure}
\begin{center}
\includegraphics[height=9cm,width=10cm]{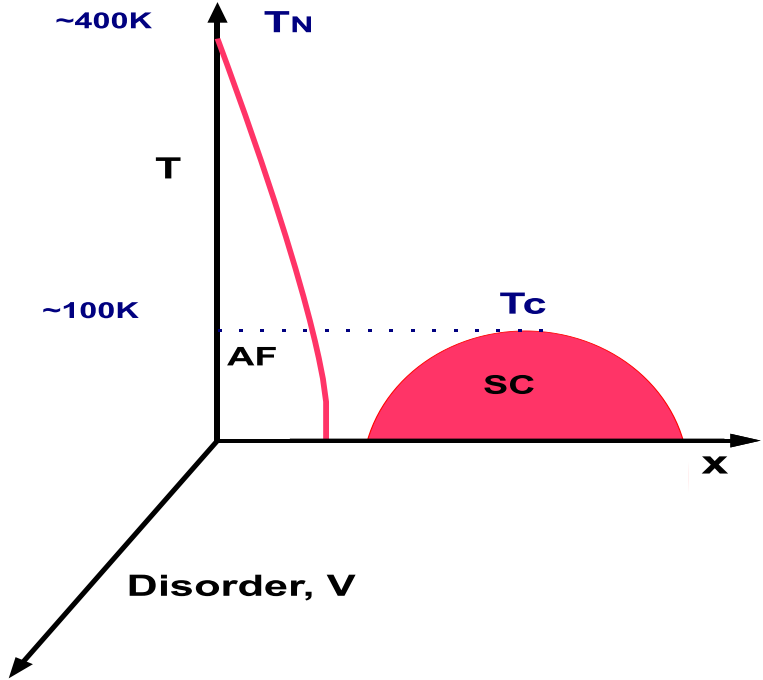}
\caption{\label{fig:sc_phase} Systematic phase diagram of the copper oxide materials.}
\end{center}
\end{figure}

The study of the Hubbard model has been stimulated after the discovery of cuprate superconductors, such as $\mathrm{YBa_2Cu_3O_{7-x}}$ where $\mathrm{x}$ is the doping concentration and $\mathrm{x}=1-n$ where $n=\frac{\mathrm{N}_{\mathrm{elec}}}{\mathrm{N}_{\mathrm{site}}}$ is the concentration of electrons \cite{bednorz1986}. A systematic phase diagram of the copper oxide materials is shown in Fig.~\ref{fig:sc_phase} where $\mathrm{x}=0$ corresponds to half-filling with 1 electron per site.
At $\mathrm{x}=0$ the system is a Mott insulator with anti-ferromagnetic order. Upon doping, the N\'{e}el temperature, $T_N$, decreases. On the other hand, $T_c$ shows a dome shaped dependence on x. This doping not only gives superconductivity but also introduces disorder.
 A realistic modeling will therefore need to consider both effects of electron correlation and disorder.

Our focus is to understand the competition between random disorder and electron-electron interaction on the Mott phase with long-range AF order in 2D. In order to characterize the phases, we calculate the temperature and frequency dependent conductivity and spin susceptibility. The calculation are done for the two-dimensional disordered Hubbard model on lattices of size up to $32\times32$ at $T=0$ and as function of temperature and disorder using a numerical Hartree-Fock self-consistent method. Our results are reported at half filling for interaction strength, $U=4t$.

The remainder of this paper is organized as follows. In Section \ref{sec:dis_model}, we give the model and the Hartree-Fock method. In Section \ref{sec:dis_method}, the self-consistent numerical method is presented. In Section \ref{sec:dis_result}, we give our results of the conductivity and spin susceptibility. This is followed by a concluding discussion and an outline of possible future research.

\section{Model \label{sec:dis_model}}
We consider the disordered strongly correlated electron system described by the 2D one-band Hubbard model with site disorder:
\begin{equation}
H=-t\sum_{
\langle ij\rangle,\sigma}\left(c_{i\sigma}^{\dag}c_{j\sigma}+c_{j\sigma}^{\dag}c_{i\sigma}\right)
+\sum_iUn_{i\uparrow}n_{i\downarrow}+\sum_{i,\sigma}(V_i-\mu)c_{i\sigma}^{\dag}c_{i\sigma}
\end{equation}
where $t$ is the near-neighbor hopping amplitude, $U$ is the on-site repulsion between electrons, $c_{i\sigma}^{\dag}$ ($c_{i\sigma}$) is the creation (destruction) operator for an electron with spin $\sigma$ on site $\vec{r}_i$, $n_{i\sigma}=c_{i\sigma}^{\dag}c_{i\sigma}$ is the number operator at site $\vec{r}_i$, and $\mu$ is the chemical potential. $V_i$ is local disorder potential at site $\vec{r}_i$ and is chosen from a uniform distribution, $[-V,V]$, where $V$ is the measure of the strength of disorder.

The Hubbard model with $U\simeq 10t$ describes the essential physics of the copper oxide, $\mathrm{CuO_2}$, perovskites upon doping. At half filling, $\langle n\rangle=1$, with no disorder, the system at $T=0$ has long-range antiferromagnetic (AF) order and is a Mott-type insulator.

The mean-field approximation is applied for the electron-electron interaction with the following prescription:
\begin{equation}
n_{i\uparrow}n_{i\downarrow}
\simeq\langle n_{i\downarrow}\rangle n_{i\uparrow}
+\langle n_{i\uparrow}\rangle n_{i\downarrow}
-\langle n_{i\downarrow}\rangle \langle n_{i\uparrow}\rangle
-c_{i\uparrow}^{\dag}c_{i\downarrow}\langle c_{i\downarrow}^{\dag}c_{i\uparrow}\rangle
-c_{i\downarrow}^{\dag}c_{i\uparrow}\langle c_{i\uparrow}^{\dag}c_{i\downarrow}\rangle
+\langle c_{i\uparrow}^{\dag}c_{i\downarrow}\rangle\langle c_{i\downarrow}^{\dag}c_{i\uparrow}\rangle .
\end{equation}
Then the effective Hamiltonian, ignoring constant terms, $\!-\!\!\sum_i\!U\{\!\langle n_{i\uparrow}\rangle\langle n_{i\downarrow}\rangle+\langle c_{i\uparrow}^{\dag}c_{i\downarrow}\rangle\langle c_{i\downarrow}^{\dag}c_{i\uparrow}\rangle\}$, becomes
\begin{equation}
H_{eff}
=-t\sum_{\langle i j\rangle,\sigma}c_{i\sigma}^{\dag}c_{j\sigma}
+\sum_{i,\sigma}(V_i-\mu)c_{i\sigma}^{\dag}c_{i\sigma}
+\sum_{i,\sigma}U\langle n_{i\bar{\sigma}}\rangle c_{i\sigma}^{\dag}c_{i\sigma}+\sum_i(h_i^-c_{i\uparrow}^{\dag}c_{i\downarrow}+h_i^+c_{i\downarrow}^{\dag}c_{i\uparrow}),
\end{equation}
where $h_i^+=-U\langle c_{i\uparrow}^{\dag}c_{i\downarrow}\rangle$, $h_i^-=-U\langle c_{i\downarrow}^{\dag}c_{i\uparrow}\rangle$, $\bar{\uparrow}=\downarrow$, and $\bar{\downarrow}=\uparrow$. In our calculation, we assume $h_i^+$ and $h_i^-$ are real so $h_i=h_i^+=h_i^-$. We therefore have 3N variational parameters, $\langle n_{i\sigma}\rangle$ for both spin species and $h_i$ at the $N$ site that must be determined self-consistently.

Explicitly, the effective Hamiltonian for $N$ lattice sites in the bases, ${|1\!\!\uparrow\rangle}, {|1\!\!\downarrow\rangle}, {|2\!\!\uparrow\rangle},  {|2\!\!\downarrow\rangle}$ $...$ ${|N\!\!\uparrow\rangle}, {|N\!\!\downarrow\rangle}$ with periodic boundary conditions is:
\begin{equation}
\left(
\begin{array}{ccccccccc}
W_{1\uparrow} & h_1 & -t & 0 & .&. &. & -t & 0 \\
h_1 & W_{1\downarrow} & 0 & -t & .&.&. &  0 & -t \\
-t & 0 & W_{2\uparrow} & h_2 & .&.&. & 0 & 0 \\
0 & -t & h_2 & W_{2\downarrow} & .&.&. & 0 & 0 \\
. & . & . & . & . &   &    & . & . \\
. & . & . & . &   & . &    & . & . \\
. & . & . & . &   &   & .  & . & . \\
-t & 0 & 0 & 0 & .&.&. & W_{N\uparrow} & h_N \\
0 & -t & 0 & 0 & .&.&. & h_N & W_{N\downarrow} \\
\end{array}
\right),
\end{equation}
where $W_{i\sigma}=V_{i}+U\langle n_{i\bar{\sigma}}\rangle-\mu$.

\section{Method \label{sec:dis_method}}
The inhomogeneous Hartree-Fock approximation for the disordered Hubbard model generates a matrix with local densities, $\langle n_{i\sigma}\rangle$, and local magnetic fields, $h_i$, which are variational parameters and must be solved self-consistently for each $\mu$ \cite{heidarian2004,heidarian2005}. For half filling, $\mu=U/2$ with no disorder but in the presence of disorder, $\mu$ is adjusted to satisfy the half filling condition. The initial input parameters for $\langle n_{i\sigma}\rangle$ and $h_i$ are determined as follows: For low disorder we start the self-consistent procedure with an AF initial condition for $\langle n_{i\sigma}\rangle$. On the other hand, for high disorder a paramagnetic state is used for the stating configuration. For both cases, $h_i$ are chosen from an AF initial condition from $-0.1t$ or $0.1t$. The initial condition for $h_i$ should not affect the results. The Broyden method is used for achieving self-consistency efficiently \cite{numerical_recipes}. Input and output fields are compared after each iteration and if the difference of the fields at all sites is less than $10^{-4}$, the self-consistentcy loop is exited. Using these final self-consistent fields, $\langle n_{i\sigma}\rangle$ and $h_i$, and eigenvalues, physical quantities of interest such as conductivity are calculated. We also test that the self-consistent values of the parameters are not dependent on the initial starting conditions for the parameters.

\section{Results \label{sec:dis_result}}



Our main effort in this part of the thesis is to calculate the frequency and temperature dependent conductivity and spin susceptibility, which are experimentally measurable and provide further characterization of these phases.

Summary of results are the following. In Region 2, the conductivity near $\omega=0$ is most enhanced and shows the Drude like behavior and the peak frequency, where the conductivity has the maximum value, is lowest. The Drude model can explain non-monotonic behavior of the dc conductivity as a function of $T$. The result of ac conductivity shows non-Fermi liquid type behavior in Region 2. The sum rule is justified except for high $T$ at low disorder. The temperature dependent behavior of spin susceptibility, $\chi$, shows the nature of moments. At high $T$, the moments are free as indicated by their Curie behavior; at low $T$, $\chi$ is strongly suppressed because electrons get paired by forming singlet on low disorder and by on-site pairing on high disorder sites.


\subsection{Frequency dependent conductivity at $T=0$}
\begin{figure}
\begin{center}
\includegraphics[height=8cm,width=10cm]{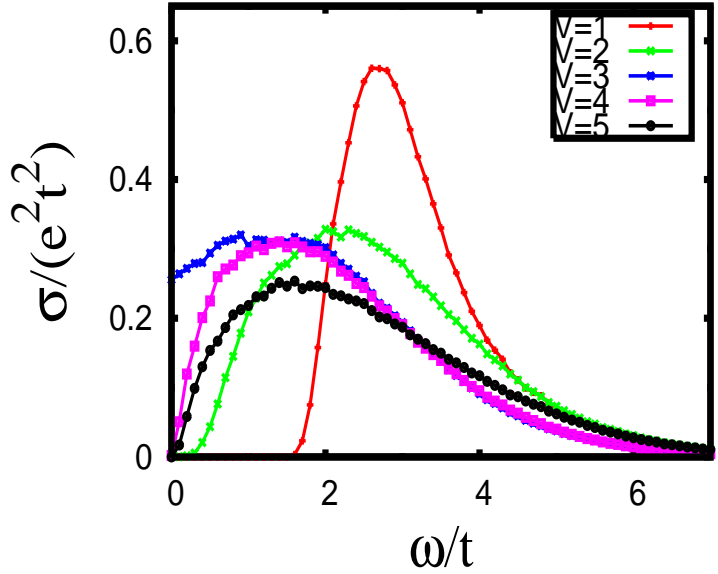}
\caption{\label{fig:fig1} The frequency dependence of the conductivity, $\sigma(\omega)$, for $V_d=1, 2, 3, 4,$ and $5$. The system size is $32\times 32$, $U=4t$, and the data is averaged over $20$ disorder realizations. The low frequency region of $\sigma (\omega)$ for $V_d=3t$ is obtained by a third order polynomial fit of the current-current correlation function, $\Lambda=\omega\sigma$. For $V_d=1t$ there is clearly a finite Mott gap at the low frequency. Thus, $\sigma(\omega)$ is suppressed at low frequency and there is a peak in the frequency when $\omega$ is around the Mott gap. With increasing disorder the spectral weight above the Mott gap is transfered to lower spectral region and the peak of $\sigma(\omega)$ moves to the low frequency. At $V_d\simeq 2.2t$, the gap in the conductivity closes because energy scale of disorder is comparable with the gap. Near $V_d=3t$, the conductivity near $\omega=0$ is most enhanced and the frequency where the conductivity has the maximum value is lowest. For $V_d=4t$ and $V_d=5t$, the dc conductivity, $\sigma(0)$, is zero due to the Anderson insulating regime. In Region 2, the peak location of $\sigma(\omega)$ moves to the low frequency as the Mott gap closes but with increasing disorder the peak location again moves to the high frequency in Region 3}
\end{center}
\end{figure}

\begin{figure}
\begin{center}
\begin{tabular}{rl}
\resizebox{70mm}{60mm}{\includegraphics{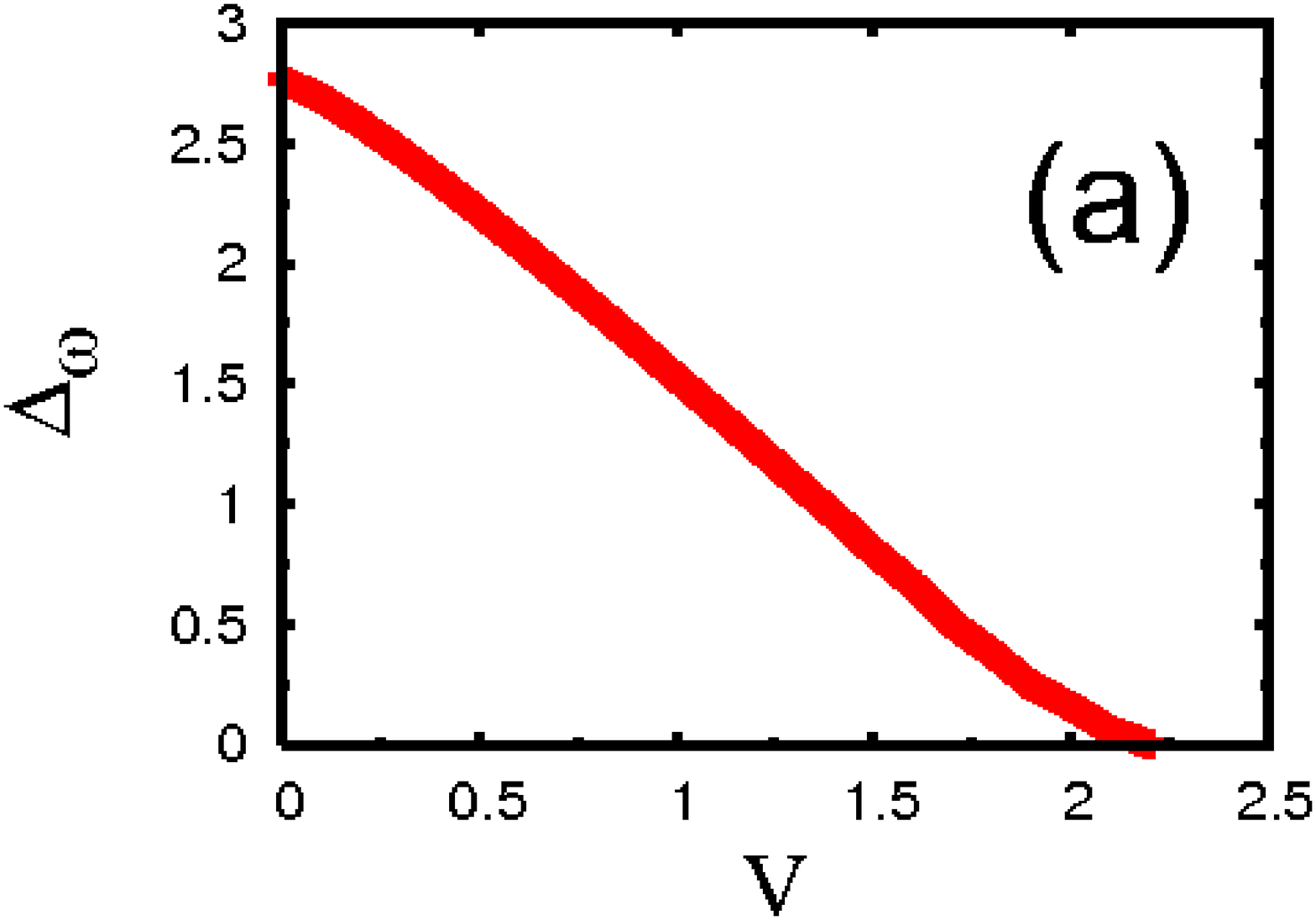}} &
\resizebox{70mm}{60mm}{\includegraphics{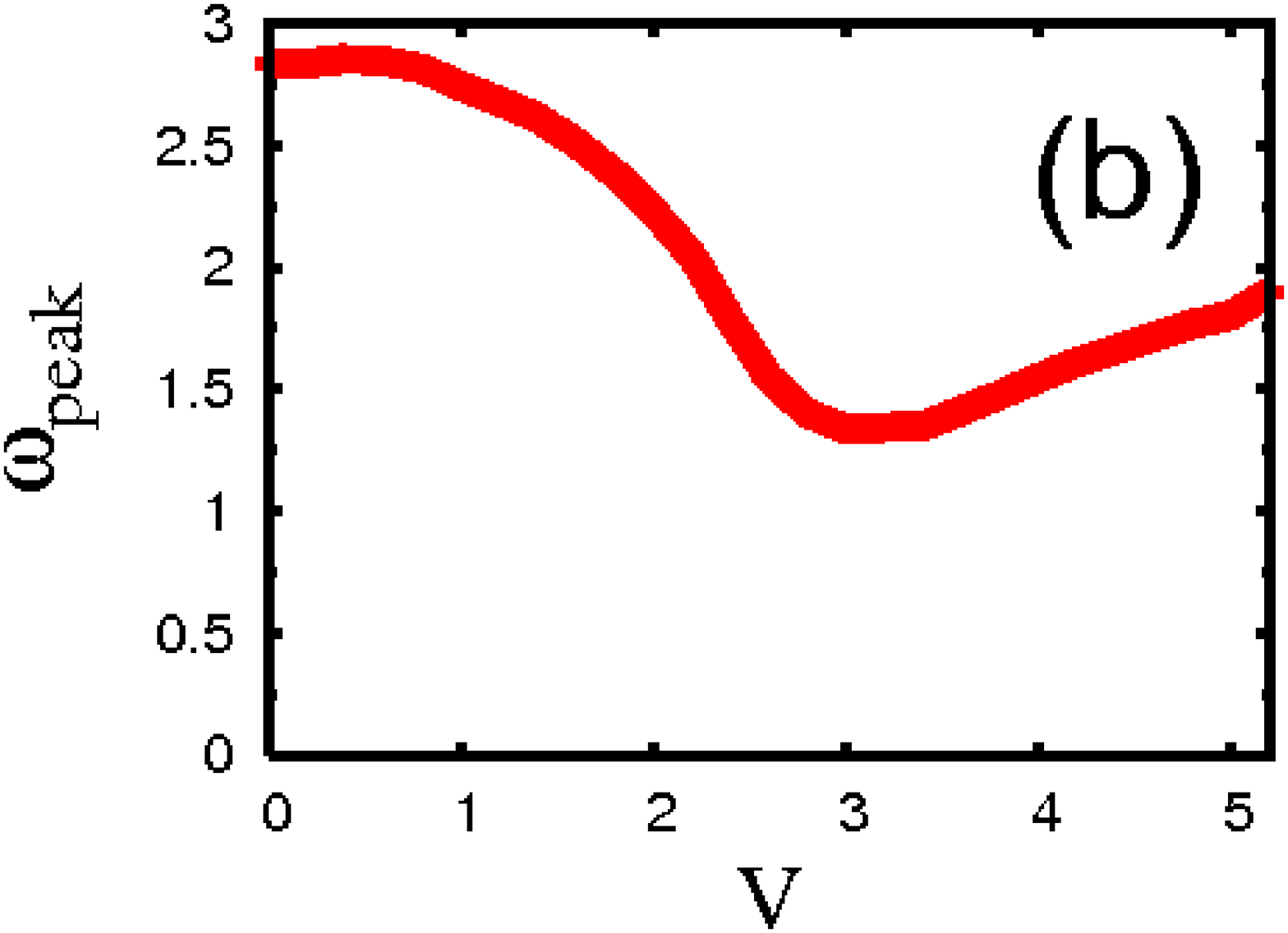}}
\end{tabular}
\caption{\label{fig:fig2} (a) The gap in $\mathrm{Re}[\sigma(\omega)]$ as a function of disorder, $V_d$. (b) $\omega_{\mathrm{peak}}$ as a function of disorder. The system size is $32\times 32$, $U=4t$, and the data is averaged over $20$ disorder realizations. $\omega_{\mathrm{peak}}$ is obtained by fitting $\mathrm{Re}[\sigma(\omega)]$ in Fig.~\ref{fig:fig1} near the peak using Lorentzian. The gap in the conductivity clearly closes at $V_d\simeq 2.2t$. $\omega_{\mathrm{peak}}$ in Region 2 is lower than that in Region 1 and 3.}
\end{center}
\end{figure}

The real part of frequency dependent conductivity, $\sigma(\omega)$, is related to the imaginary part of the current-current correlation function \cite{comment_sigma},
\begin{equation}
\mathrm{Re}\left[\sigma(\omega)\right]=\frac{\mathrm{Im}\Lambda(\omega)}{\omega},
\end{equation}
where $\Lambda(\omega)$ is the Fourier transform of $\Lambda(\tau)=\langle j(\tau)j(0)\rangle$ and $j$ is the paramagnetic current density operator. Fig.~\ref{fig:fig1} shows $\sigma(\omega)$ for various disorder strengths and Fig.~\ref{fig:fig2} shows the gap of the conductivity as a function of disorder and the value of $\omega$ where the conductivity has a peak as a function of disorder.
For low disorder, Region 1, there is a gap in the conductivity which is twice the spectral gap seen in the DOS and a strong peak in the absorption at $\omega\simeq U$. At intermediate disorder, the gap in $\sigma(\omega)$ closes at $\omega=0$ in Fig.~\ref{fig:fig2} (a) but the peak in $\mathrm{Re}[\sigma(\omega)]$ persists. As seen in Fig.~\ref{fig:fig2} (b), the peak moves to lower values of $\omega$ as disorder increases in Region 2. In Region 3, the peak starts moving out to higher frequencies but the conductivity at high $T$ is still steeper than $\mathrm{Re}[\sigma(\omega)]\sim 1/\omega^{2}$. With increasing disorder, the spectral weight above the Mott gap is transfered to lower spectral region, enhancing the absorption at a low frequency. At $V_d=2.2t$, the gap of the conductivity closes because energy scale of disorder is comparable with the gap. Near $V_d=3t$, $\mathrm{Re}[\sigma(\omega=0)]$ is most enhanced and $\omega_{peak}$, frequency where the conductivity has the maximum value, is lowest. The optical conductivity at $V_d=3t$ shows a Drude-like behavior. As disorder increases, this Drude-like behavior disappears. The origin of this Drude-like behavior at $V_d=3t$ may be the result of the screening of the strongly disordered paramagnetic sites with 2 electrons on a site. The disordered potentials are screened and an electron feels the following density dependent effective potentials.
\begin{eqnarray*}
\tilde{V}_{i\uparrow} &=& V_i-\mu+U\langle n_{i\downarrow}\rangle \\
\tilde{V}_{i\downarrow} &=& V_i-\mu+U\langle n_{i\uparrow}\rangle
\end{eqnarray*}
where the effect to an electron comes from the density with opposite spin. Ignoring spin-polarization, this effective potential becomes
\begin{equation}
\tilde{V}=\frac{1}{2}(\tilde{V}_{i\uparrow}+\tilde{V}_{i\downarrow}) = V_i-\mu+\frac{U}{2}\left(\langle n_{i\uparrow}\rangle+\langle n_{i\downarrow}\rangle\right).
\end{equation}
For $V_d=4t$ and $V_d=5t$, although the Mott gap is completely filled, the conductivity is suppressed due to the Anderson insulating regime. This insulating behavior continues and is enhanced more as disorder increases. It is difficult to fit the conductivity at high frequency by the form $(1/\omega^\alpha)$ and $\sigma(\omega)$ for all case has a much faster decay with $\alpha=2$, which is a conventional Fermi liquid result.


\subsection{DC conductivity}
\begin{figure}
\begin{center}
\includegraphics[height=8cm,width=10cm]{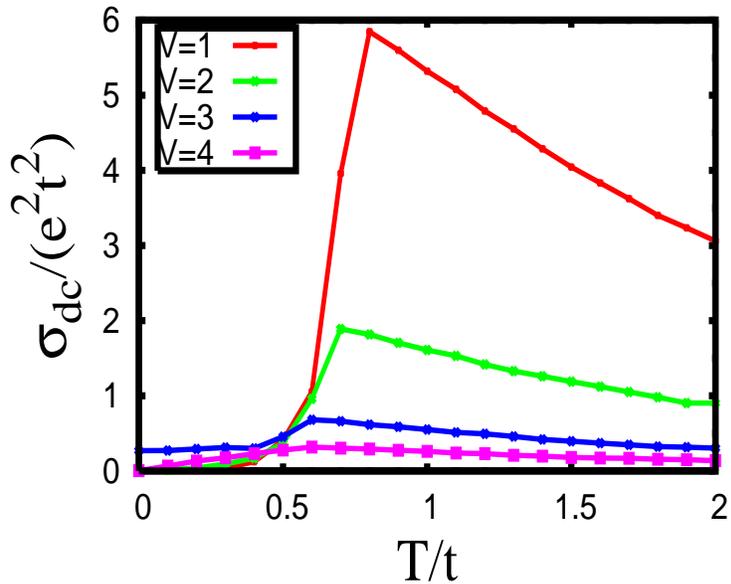}
\caption{\label{fig:fig3} Temperature dependence of the dc conductivity, $\sigma_{dc}(\omega)$, for $V_d=1,2,3$, and $4$. The system size is $16\times 16$, $U=4t$, and the data is averaged over $20$ disorder realizations. The dc conductivity at low $T$ for $V_d=3t$ is obtained by a third order polynomial fit of the current-current correlation function, $\Lambda=\omega\sigma$. At low $T$, the dc conductivity in Region 1 is zero because of the Mott gap, the system with $V_d=3t$ is more conducting than $V_d=1t$ due the close of the Mott gap, and with increasing disorder, $\sigma_{dc}$ is suppressed because of an Anderson localization of electrons. On the other hand, at high $T$, $\sigma_{dc}$ in Region 1 is higher than the others because energy scale of temperature is comparable with the Mott gap and as disorder increases, $\sigma_{dc}$ decreases due to the increase of transport scattering of electrons and disorder. It is demonstrated that disorder can drive a metallic phase in the system. At low temperature, the dc conductivity in Region 1 and 2 decreases as lowering temperature, signifying an insulating phase. However, the dc conductivity for $V_d=3t$ has been most enhanced and slope of the conductivity, $\sigma_{dc}/dT$, has no slope.}
\end{center}
\end{figure}

DC conductivity is obtained by
\begin{equation}
\sigma_{dc}=\mathrm{lim}_{\omega\rightarrow 0}\frac{\mathrm{Im}\Lambda(\omega)}{\omega}.
\end{equation}
Fig.~\ref{fig:fig3} shows the temperature dependence of $\sigma_{dc}$ for various $V$.
In the Mott insulator (e.g. $V_d=1t$ or $V_d=2t$), $\sigma_{dc}$ is exponentially small at low $T$, it peaks at a temperature of order $U$, and at higher $T$, $\sigma_{dc}$ decreases with $T$. In the intermediate disorder regime the system with $V_d=3t$ is more conducting than $V_d=1t$ at low $T$ but at high $T$ it is the opposite way. Increasing the site disorder further drives the system from a metallic to an Anderson insulating phase and the low $T$, $\sigma_{dc}$ drops again. The dc conductivity at low $T$ and at low disorder e.g. $0<V<2$ is zero because of the Mott gap, $\sigma_{dc}$ at moderate disorder e.g. $2<V<3.5$, has a finite value, and then again $\sigma_{dc}$ is zero due to the localization of electron on the strongly disordered site. The dc conductivity shows non-monotonic behavior as a function of $T$ which can be understood using the Drude model:
\begin{equation}
\sigma_{dc}^{\mathrm{Drude}}= \frac{ne^2\tau}{m},
\end{equation}
where $\tau$ is the relaxation time and $n$ is the density of electrons. At high temperature $\sigma_{dc}$ is dominated by decreasing $\tau$. For increasing disorder, $\tau$ decreases because of enhanced scattering between electrons and the disorder potential. At low temperature $\sigma_{dc}$ is dominated by the temperature dependence of $n$. The number of electrons need to be excited beyond the Mott gap so the density is proportional to $e^{-\Delta_{Mott}/T}$. Therefore, the conductivity at low disorder increases rapidly with increasing temperature. For both low and high disorder, $\sigma_{dc}$ vanishes as $T\rightarrow 0$, signifying an insulating phase, either a Mott insulator or an Anderson insulator. However, as $T\rightarrow 0$ the dc conductivity for $V_d=3t$ is unusual; it is finite with negligible $T$-dependence at low $T$.

\subsection{Frequency dependence of the conductivity for finite temperature}
\begin{figure}
\begin{center}
\begin{tabular}{rl}
\resizebox{72mm}{70mm}{\includegraphics{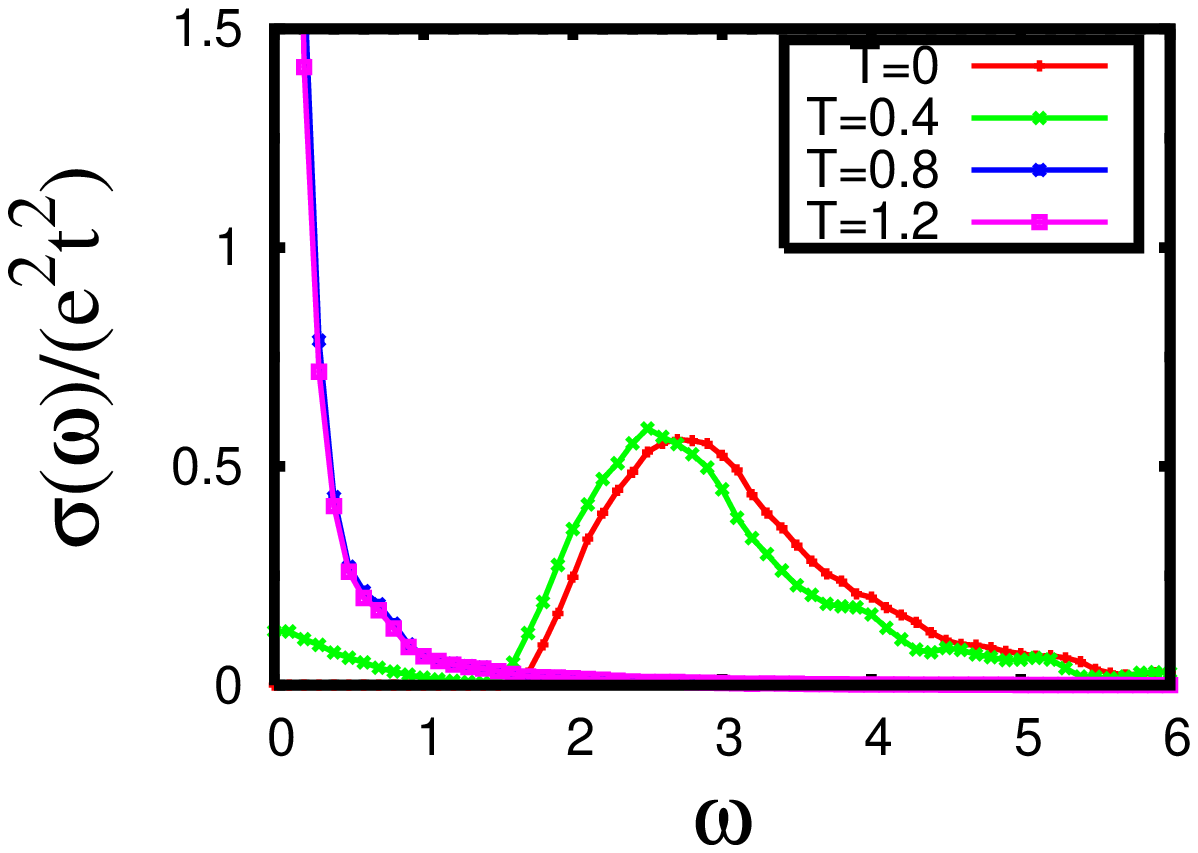}} &
\resizebox{72mm}{70mm}{\includegraphics{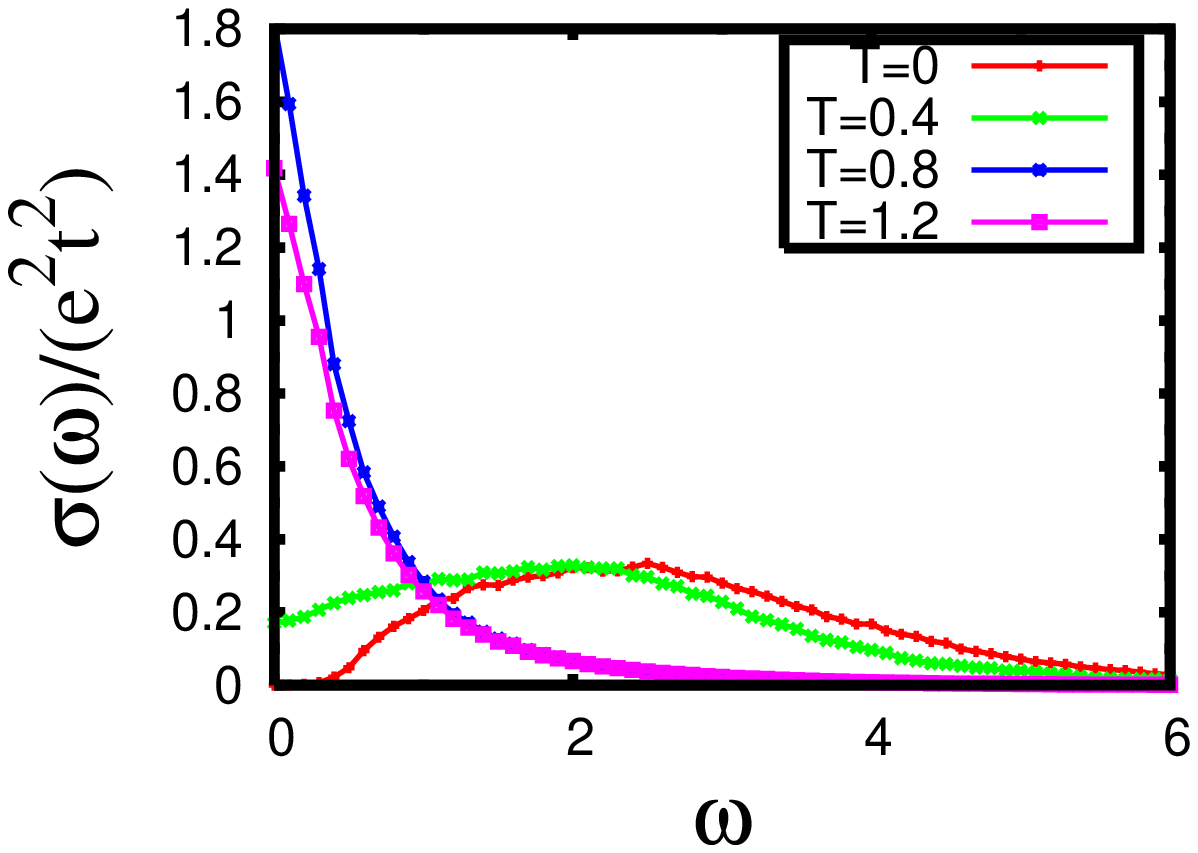}} \\
\resizebox{72mm}{70mm}{\includegraphics{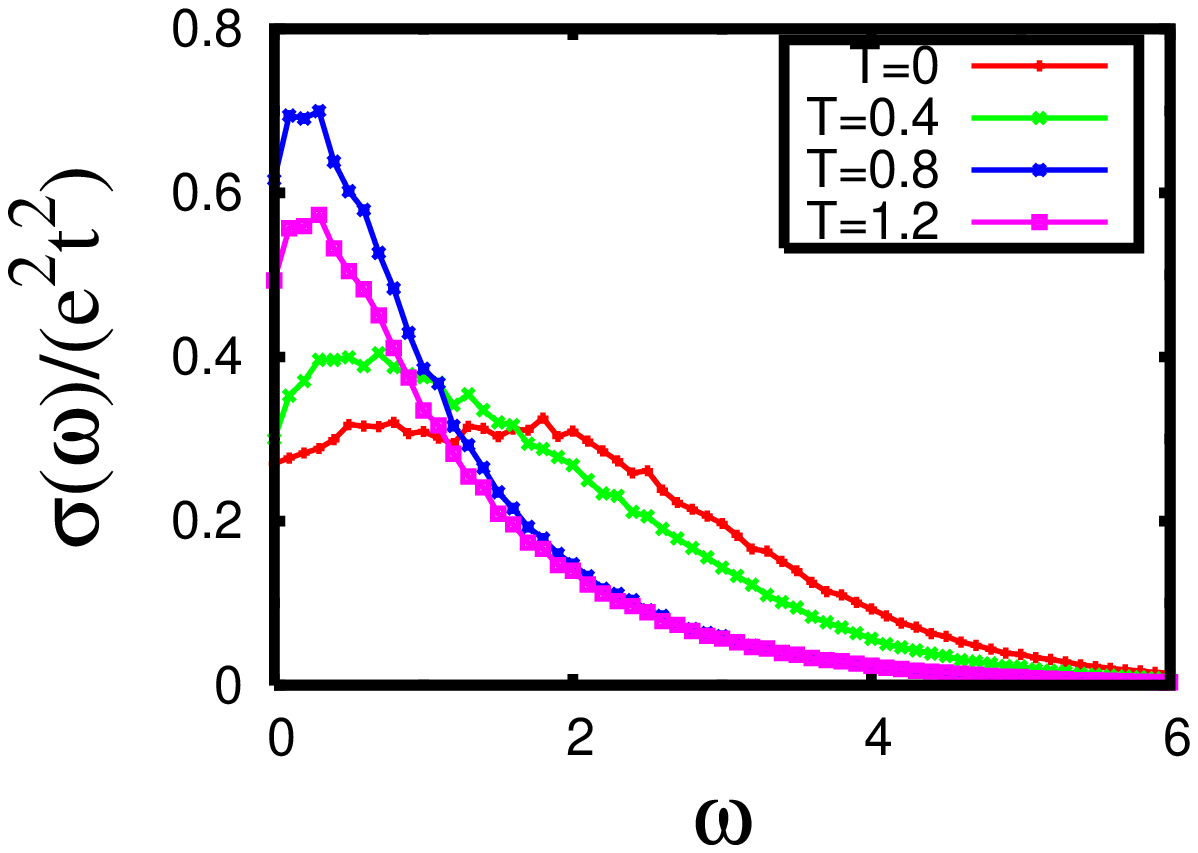}} &
\resizebox{72mm}{70mm}{\includegraphics{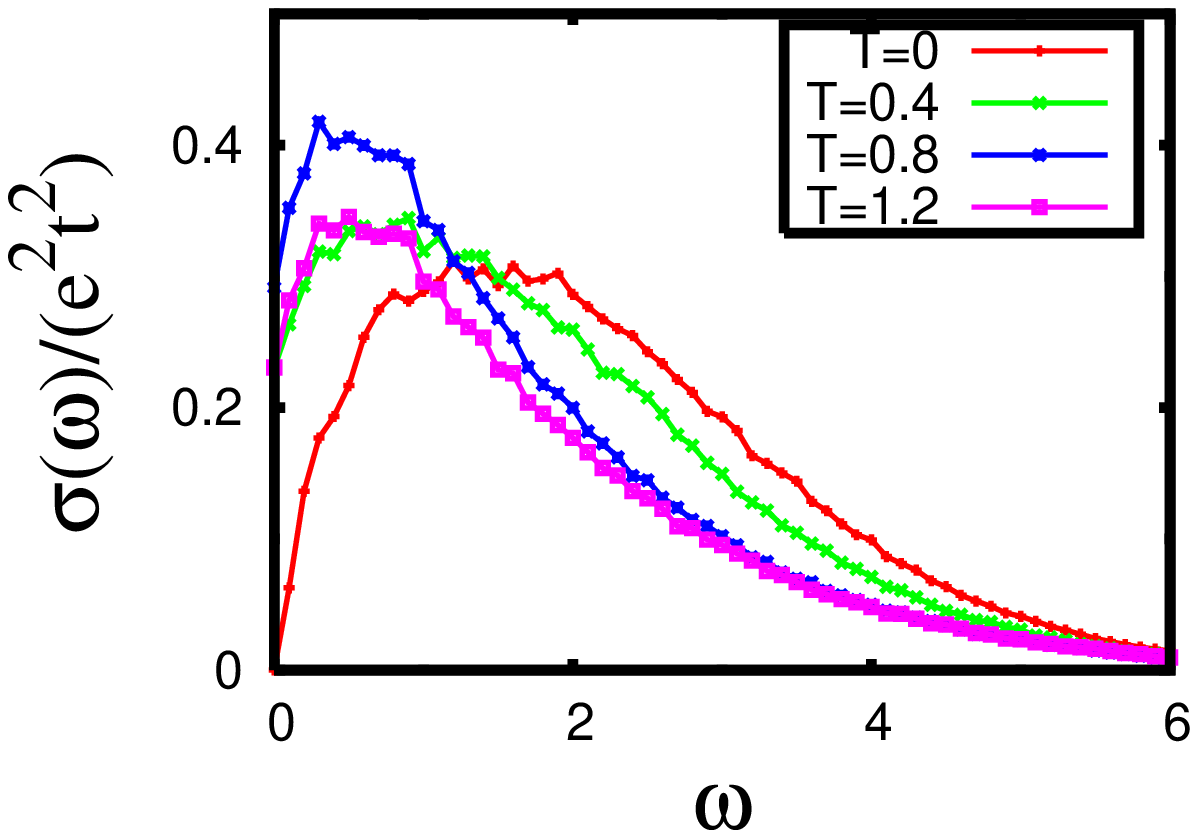}} \\
\end{tabular}
\caption{\label{fig:fig4} The frequency dependence of the conductivity for $V_d=1$ (left top), $V_d=2$ (right top), $V_d=3$ (left bottom) and $V_d=4$ (right bottom) at different temperatures. The system size is $16\times 16$, $U=4t$, and the data is averaged over $20$ disorder realizations. The low frequency region of $\sigma (\omega)$ for $V_d=3t$ is obtained by a third order polynomial fit of the current-current correlation function, $\Lambda=\omega\sigma$. For low disorder at low $T$ there is the Mott gap for low frequency and with increasing $T$, the peak in the conductivity moves to the weight at $\omega=0$. With increasing disorder from Region 1 to Region 2, the spectral weight above the Mott gap is transferred to lower spectral region at low $T$ and $\sigma(\omega)$ shows the Drude like behavior for all $T$. With increasing disorder from Region 2 to Region 3, this Drude like behavior is less clear. The Drude like behavior shows non-monotonic behavior as a function of $T$. The conductivity weight in low frequency increases with increasing $T$ up to $T=0.8$ but with further increasing $T$ it decreases.}
\end{center}
\end{figure}

Fig.~\ref{fig:fig4} shows the frequency dependence of the conductivity, $\mathrm{Re}[\sigma(\omega)]$, for various disorder strengths.
Region 1: Mott Insulator at low disorder (V=1) shows a finit gap in the spectrum and a peak in $\mathrm{Re}(\omega)$ at a characteristic frequency $\omega_{\mathrm{peak}}\approx U$. With increasing $T$, $\omega_{\mathrm{peak}}$ moves to lower $\omega$ and the weight at $\mathrm{Re}\sigma (\omega\approx 0)$ up. At $T\approx t$ the weight at intermediate frequencies $\omega_{\mathrm{peak}}$ vanishes and gets concentrated in a Drude-like peak around $\omega\approx 0$. Region 2: Metalic regime at intermediate disorder (V=2,3). $\mathrm{Re}\sigma(\omega)$ shows that the Mott gap vanished at low $T$. As $T$ increases weight buids up at $\omega\approx 0$ and has a finite intarcept, $\sigma_{dc}$. At higher $T$, $\sigma(\omega)$ at $\omega\approx 0 $ develops and shows Drude-like behavior. Region 3: Insulating regime at high disorder (V=4), $\sigma_{dc}$ continues to be zero even at higher temperature. $\omega_{\mathrm{peak}}$ shifts to lower $\omega$ with increasing $T$.

For low disorder at low $T$ there is a Mott gap but with increasing $T$ this gap closes and $\sigma(\omega=0)$ has large weight. In Region 2, for all $T$, the Drude like conductivity is seen and in Region 3, this behavior is less obvious. AC conductivity from the Drude theory is
\begin{equation}
\sigma(\omega) = \frac{\sigma_0}{1-i\omega\tau},
\end{equation}
where $\sigma_0=\frac{ne^2\tau}{m}$. The real part is
\begin{equation}
\sigma^{\mathrm{Drude}} (\omega) = \frac{\sigma_0}{1+\omega^2\tau^2}.
\end{equation}
At high frequency, $\sigma^{\mathrm{Drude}} (\omega)= ne^2/(m\omega^2\tau)$. It is difficult to fit the conductivity at high frequency by the form $(1/\omega^\alpha)$ because $\sigma(\omega)$ for all cases has a much faster decay with $\alpha=2$. The sudden increase of the conductivity at low $\omega$ with increasing $T$ for $V_d=1$ and $V_d=2$ may be explained by the exponential increase of the DOS at the Mott insulator as a function of temperature.

\subsection{Sum rule for the conductivity}
\begin{figure}
\begin{tabular}{cc}
\resizebox{72mm}{70mm}{\includegraphics{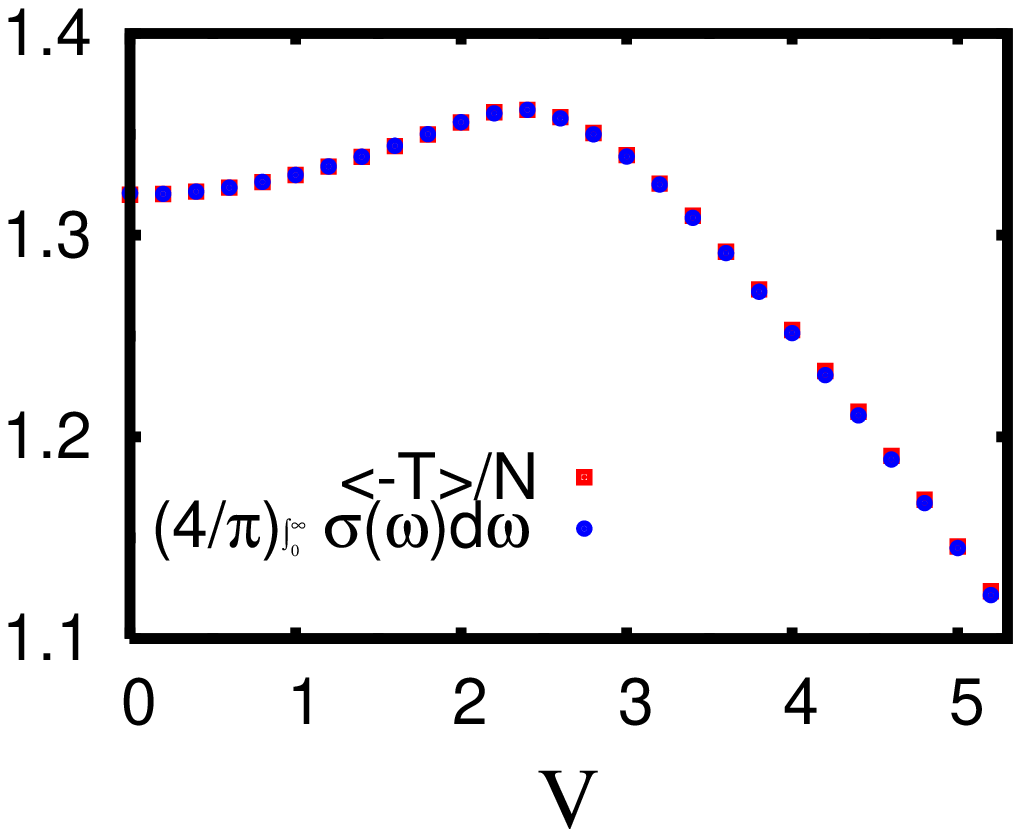}} &
\resizebox{72mm}{70mm}{\includegraphics{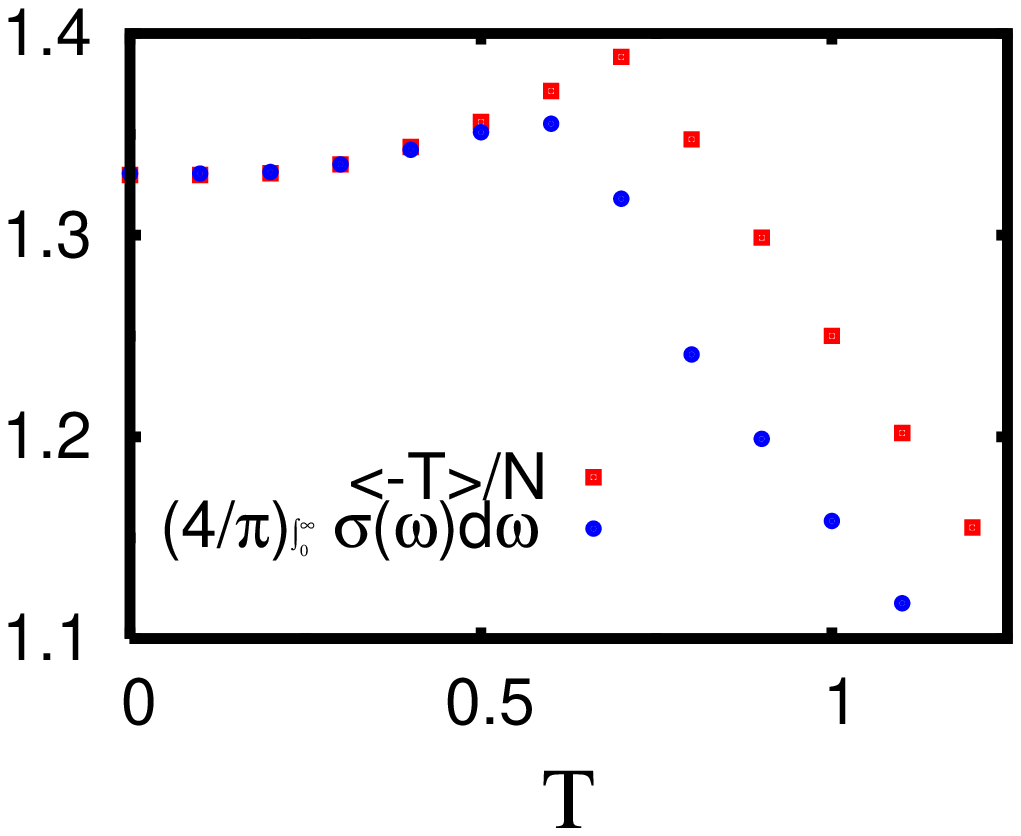}} \\
\resizebox{72mm}{70mm}{\includegraphics{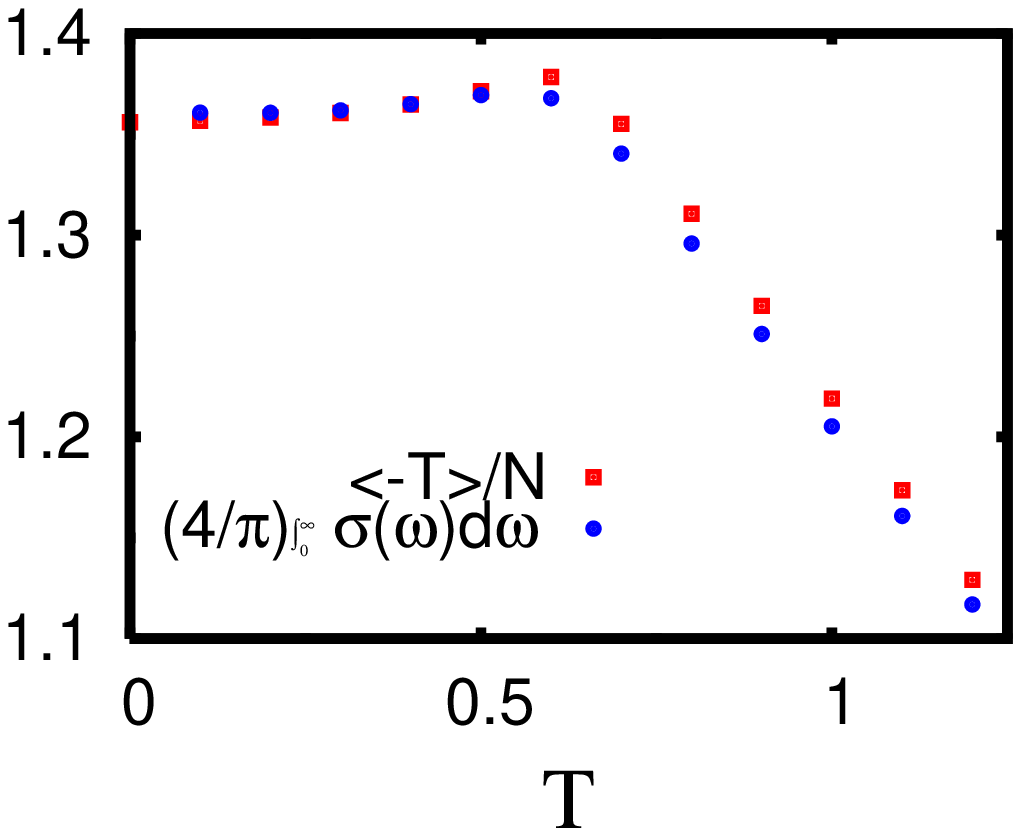}} &
\resizebox{72mm}{70mm}{\includegraphics{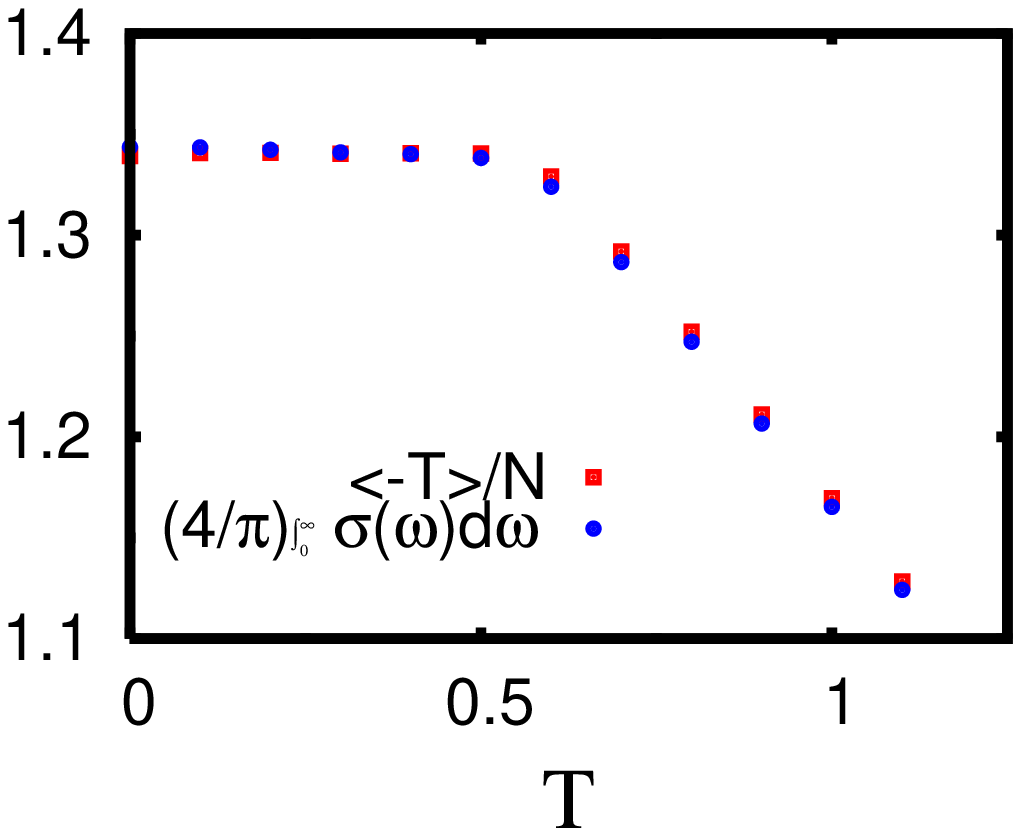}} \\
\end{tabular}
\caption{\label{fig:fig5} Comparison of $\frac{\langle -\hat{T}\rangle}{N}$ (red square) where $\hat{T}=t\sum_{\langle i j\rangle,\sigma}c_{i\sigma}^{\dag}c_{j\sigma}$ and $\frac{4}{\pi e^2}\int_0^{\infty}\sigma (\omega)d\omega $ (blue dot) as (a) a function of disorder for $T=0$ or a function of a temperature for (b) $V_d=1$, (c) $V_d=2$, and (d) $V_d=3$. The system size is $16\times 16$, $U=4t$, and the data is averaged over $20$ disorder realizations. The agreement between the blur and red dots indicates that the f-sum rule is obeyed. The disagreement in (b) at low disorder (V=1) comes from the difficulty of calculating the Drude weight in the presence of singular $\delta$-function at $\omega=0$.}
\end{figure}

As a test of our numerical results, we check the consistency of the f-sum rule \cite{maldague1977}.
The frequency integral of $\mathrm{Re}[\sigma(\omega)]$ is related to the total kinetic energy: 
\begin{equation}
\frac{\langle -\hat{T}\rangle}{N}=\frac{4}{\pi e^2}\int_0^{\infty}\mathrm{Re}[\sigma (\omega)]d\omega, \label{eq:sumrule}
\end{equation}
where $\hat{T}=t\sum_{\langle i j\rangle,\sigma}c_{i\sigma}^{\dag}c_{j\sigma}$. We calculate and compare the left-hand side and the right-hand side of Eq.~(\ref{eq:sumrule}), independently in Fig.~\ref{fig:fig5}.
These values are non-monotonic as functions of both $T$ and disorder. The average kinetic energy, $\langle\hat{T}\rangle$, increases with temperature at low $T$ but decreases with $T$ for higher temperature. Similarly, with disorder, the kinetic energy increases for low disorder around the Mott insulator but at high disorder the kinetic energy decreases with increasing disorder because the mobility of electrons is decreased at strongly disordered sites. At low temperature, the sum rule is in good agreement. At finite temperature the agreement in the sum rule improves at higher disorder. The disagreement at high temperature for $V_d=1$ comes from the difficulty of calculating exactly the Drude weight, which has a singular behavior at $\omega=0$.

\subsection{Uniform static spin susceptibility}
\begin{figure}
\begin{center}
\begin{tabular}{cc}
\resizebox{72mm}{70mm}{\includegraphics{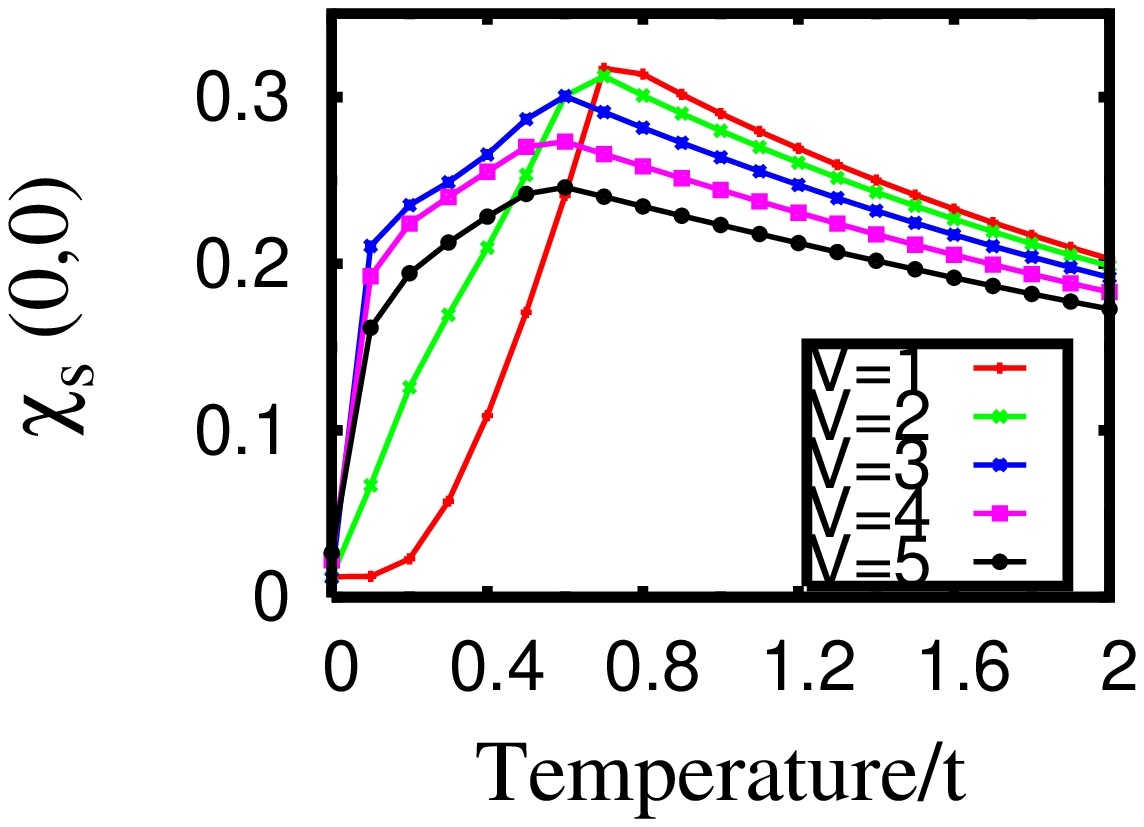}} &
\resizebox{72mm}{70mm}{\includegraphics{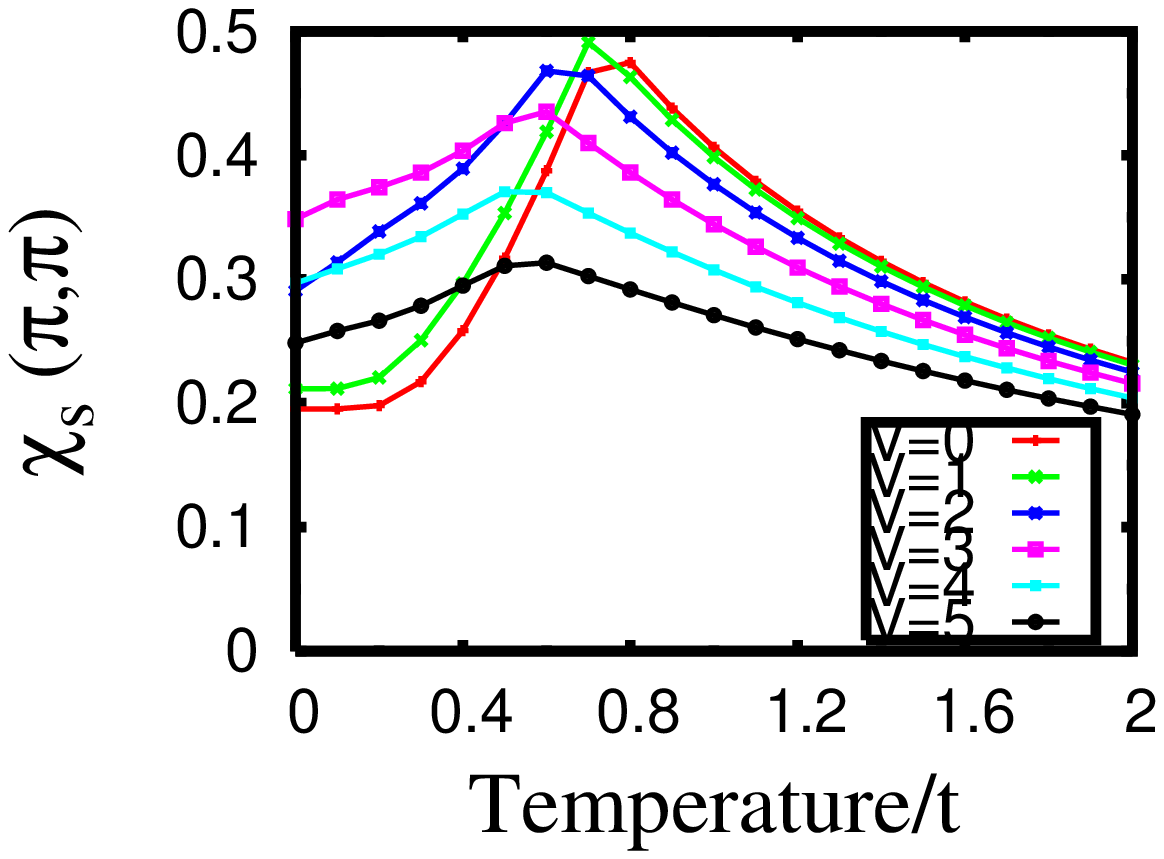}} \\
\resizebox{72mm}{70mm}{\includegraphics{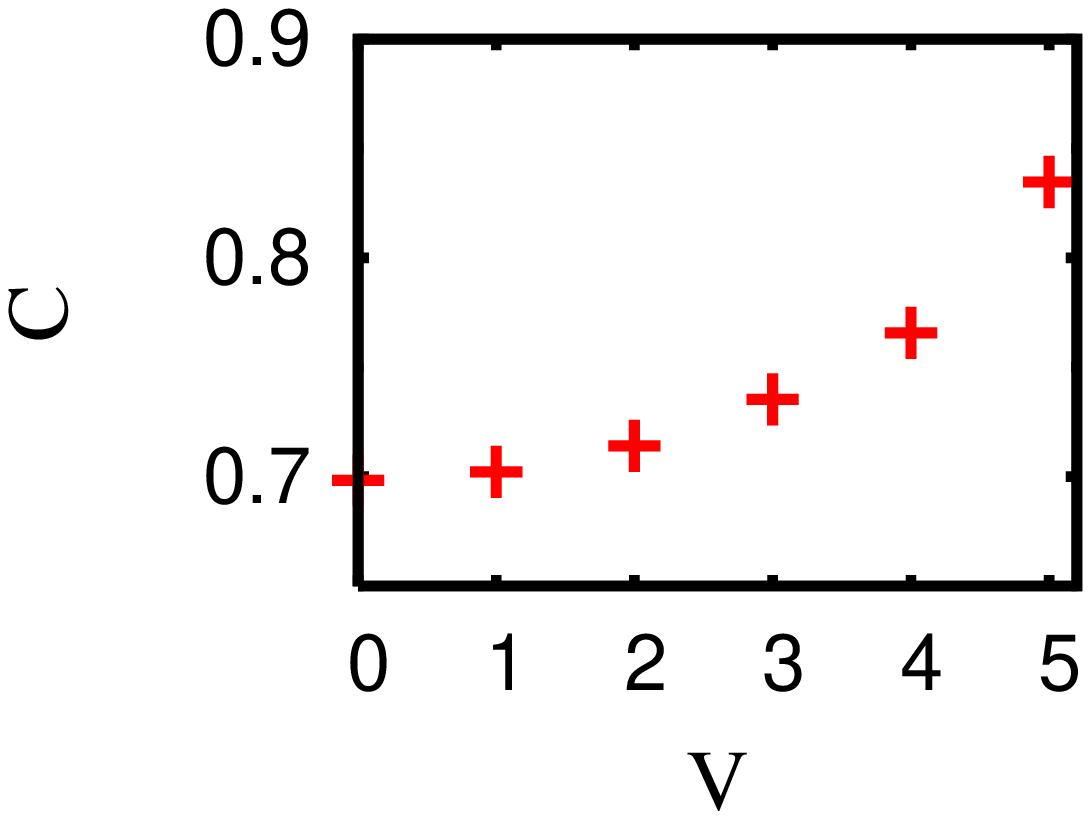}} &
\resizebox{72mm}{70mm}{\includegraphics{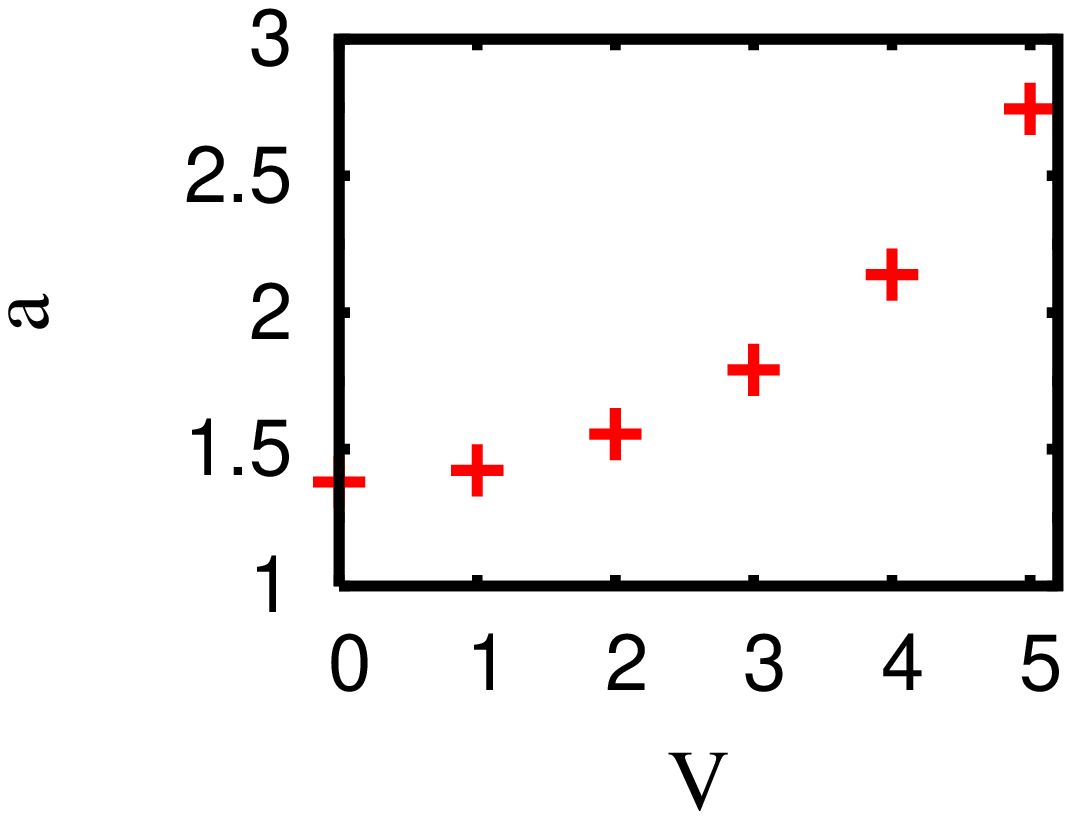}} \\
\end{tabular}
\caption{\label{fig:fig6} $\chi(0,0)$ and  $\chi(\pi,\pi)$ as a function of a temperature for $V_d=0,1,2,3,4$, and $5$. The system size is $16\times 16$, $U=4t$, and the data is averaged over $20$ disorder realizations. The tails of $\chi(0,0)$ are fitted by the function $\frac{C}{T+a}$ where $C$ is the measure of local moments and $a$ is the measure of the strength of paramagnetism. These $C$ and $a$ are plotted as a function of the disorder, V. At low temperature, $\chi(0,0)$ and $\chi(\pi,\pi)$ for low disorder are suppressed and these are most enhanced in Region 2, and with increasing disorder further, $\chi(0,0)$ and $\chi(\pi,\pi)$ decrease. On the other hand, at high temperature, increasing disorder, $\chi(0,0)$ and $\chi(\pi,\pi)$ monotonically decrease.}
\end{center}
\end{figure}

The frequency-dependent spin susceptibility is defined by
\begin{equation}
\chi^{zz}(\vec{q},\omega)=\int_0^{\beta}e^{i\omega\tau}\chi^{zz}(\vec{q},\tau)d\tau,
\end{equation}
where
\begin{equation}
\chi^{zz}(\vec{q},\tau)=\langle S_z(\vec{q},\tau)S_z(-\vec{q},0)\rangle
=\frac{1}{N}\sum_{i,j}e^{i\vec{q}\cdot(\vec{i}-\vec{j})}\langle S_i^z(\tau)S_j^z(0)\rangle
\end{equation}
and $S_z$ is $z$ component of the spin operator. Fig.~\ref{fig:fig6} shows the uniform spin susceptibility, $\chi(0,0)$, and staggered spin susceptibility, $\chi(\pi,\pi)$, for various disorder strengths as a function of $T$.
At low temperature, for low disorder, $\chi(0,0)$ vanishes indicating a spin gap. $\chi(\pi,\pi)$ is finite due to long-range AF order in the system. As $T$ increases $\chi(0,0)$ and $\chi(\pi,\pi)$ increase, reach a maximum at a characteristic temperature and decrease at higher temperatures. The spin susceptibilities are most enhanced at intermediate disorder, ($V_d=3t$).

Also, the fitted parameters, $C$ and $a$, where $\frac{C}{T+a}$ is used for the fitting the tail of $\chi(0,0)$, are considered to see local moments and the strength of paramagnetism. On the other hand, the Curie susceptibility of the spin at $k_BT\gg g\mu_BH$ for localized electrons is
\begin{equation}
\chi^{\mathrm{Curie}}=\frac{n(g\mu_B)^2S(S+1)}{3k_BT},
\end{equation}
where $n$ is the density, $g$ is 2, and $S$ is a local moment.

At high $T$ increasing disorder, the number of on-site coupling of up and down spin electrons increases because electrons can occupy the low disordered sites and not occupy the strongly disordered sites. These coupled sites and unoccupied sites have no contribution on local moments, so $\chi$ is more suppressed with increasing disorder at high $T$. The plot of $C$ justify this description. These are increased with increasing disorder as expected. On the other hand, at low $T$, we consider the Pauli paramagnetic susceptibility for free electrons:
\begin{equation}
\chi^{\mathrm{Pauli}}=\mu_B^2D(\epsilon_F),
\end{equation}
where $\mu_B$ is the Bohr magneton and $\epsilon_F$ is the DOS at the Fermi level. For low disorder, there is a spin gap because DOS$(\epsilon_F)=0$. Increasing disorder, $\chi$ is enhanced due to the increase of available states. Increasing disorder further (e.g. $V_d=4$ or $V_d=5$), this enhancement disappears because of the suppressed local moments at strong disordered sites.



\section{Conclusion}
In summary, we have calculated $\mathrm{Re}[\sigma(\omega,T)]$, the frequency and temperature dependence of conductivity, for the two-dimensional Hubbard model with disorder at half filling using an inhomogeneous self-consistent Hartree-Fock numerical method. There is a gap in $\mathrm{Re}[\sigma(\omega)]$ at low disorder. The characteristic temperature dependence of the conductivity in the Mott insulator shows $\frac{d\sigma_{dc}}{dT}>0$ for low $T$ and $\frac{d\sigma_{dc}}{dT}<0$ at high $T$ arising primarily from a carrier dominated regime at low $T$ to a scattering dominated regime at high $T$. Upon increasing disorder, the spectral weight around the Mott gap is shifted to low frequencies and $\sigma_{dc}$ grows with disorder. $\sigma_{dc}$ is enhanced at intermediate disorder with $\sigma_{dc}/dT\simeq 0$ for low temperature. However, $\sigma_{dc}$ is zero in the Mott insulator and in the strongly localized regime.

Moreover, we have calculated $\chi(\vec{q}=0,\omega,T)$, uniform spin susceptibility, and $\chi_s(\vec{q}=(\pi,\pi),\omega,T)$, staggered spin susceptibility. From the results of uniform static spin susceptibility, we considered the spin formation in each Region.

We have investigated the effect of an inhomogeneous magnetic field, $h_i = -U\langle c_{i\uparrow}^{\dag}c_{i\downarrow}\rangle$ considering the case with $h_i=0$ at all sites. We could have not seen any significant difference between two cases, $h_i=0$ for all sites and $h_i\ne 0$. Thus, the inhomogeneous field can only select the z direction on the 2D system and physical quantities that do not depend on z direction are not depending on $h_i$.

It may be interesting to contrast the behavior of the conductivity for the positive-U and negative-U disordered Hubbard model. Moreover, away from half filling is interested.


\begin{thebibliography}{99}
\bibitem{comment_sigma}
The conductivity is a complex number but we only consider the real part of the conductivity in this paper so $\sigma(\omega)=\mathrm{Re}[\sigma(\omega)]$ is used unless otherwise mentioned.
\end{thebibliography}
\end{document}